\documentclass[aps,prb,preprint,groupedaddress,showpacs,amsmath,amssymb]{revtex4}

\usepackage{graphicx}
\usepackage{dcolumn}
\usepackage{bm}

\begin{document}

\title{Anisotropic thermal expansion of Fe$_{1.06}$Te and FeTe$_{0.5}$Se$_{0.5}$ single crystals}

\author{S. L. Bud'ko}
\author{P. C. Canfield}
\affiliation{Ames Laboratory US DOE and Department of Physics and Astronomy, Iowa State University, Ames, IA 50011, USA}
\author{A. S. Sefat}
\author{B. C. Sales}
\author{M. A. McGuire}
\author{D. Mandrus}
\affiliation{Materials Science and Technology Division, Oak Ridge National Laboratory, Oak Ridge, TN 37831, USA}

\date{\today}

\begin{abstract}
Heat capacity and anisotropic thermal expansion was measured for Fe$_{1.06}$Te and FeTe$_{0.5}$Se$_{0.5}$ single crystals. Previously reported phase transitions are clearly seen in both measurements. In both cases the thermal expansion is anisotropic. The uniaxial pressure derivatives of the superconducting transition temperature in FeTe$_{0.5}$Se$_{0.5}$ inferred from the Ehrenfest relation have opposite signs for in-plane and $c$-axis pressures. Whereas the Gr\"uneisen parameters for both materials are similar and only weakly temperature-dependent above $\sim 80$ K, at low temperatures (in the magnetically ordered phase) the magnetic contribution to the Gr\"uneisen parameter in Fe$_{1.06}$Te is significantly larger than electron and phonon contributions combined.

\end{abstract}


\maketitle

The discovery of superconductivity in F-doped LaFeAsO \cite{kam08a} and K-doped BaFe$_2$As$_2$ \cite{rot08a} compounds caused an increased interest in studies of the materials containing Fe-As layers as a structural unit. More recently superconductivity was reported in two other structural families that have iron-pnictogen, or iron-chalcogen layers in their structure, Li$_{1-x}$FeAs \cite{wan08a} and FeSe$_{1-x}$. \cite{hsu08a} For the latter material, an enhancement of superconducting transition temperature, $T_c$, was observed upon substitution of S or Te for Se. \cite{miz08a} Recently, large single crystals of the Fe$_{1+y}$Te$_x$Se$_{1-x}$ were grown and explored. \cite{sal09a}

Thermal expansion is professed to be uniquely sensitive to magnetic, structural and superconducting transitions \cite{kro98a}. Anisotropic thermal expansion measurements in Ba(Fe$_{1-x}$Co$_x$)$_2$As$_2$ [\onlinecite{bud09a,har09a,luz09a}] have been instrumental in inferring unusually large, anisotropic, uniaxial pressure derivatives of superconducting transition temperature, $T_c$, in these compounds.  Gr\"uneisen parameter analysis on the other hand is frequently used for comparative, thermodynamic studies of related materials. \cite{bar99a} To gain more understanding about the members of the Fe$_{1+y}$Te$_x$Se$_{1-x}$ family, in this work we present the measurements of heat capacity and anisotropic thermal expansion on its two members: non-superconducting, parent compound, Fe$_{1.06}$Te, and, close to optimally doped, superconducting, FeTe$_{0.5}$Se$_{0.5}$.
\\

Single crystals of Fe$_{1.06}$Te, and FeTe$_{0.5}$Se$_{0.5}$ were grown by Bridgeman technique. Detailed description of the crystal growth procedure and compositional analysis for these samples can be found elsewhere. \cite{sal09a} The heat capacity of the samples was measured using a hybrid adiabatic relaxation technique of the heat capacity option in a Quantum Design, PPMS-14 instrument. Thermal expansion data were obtained using a capacitive dilatometer constructed of OFHC copper; a detailed description of the dilatometer is presented elsewhere \cite{sch06a}. The dilatometer was mounted in a Quantum Design PPMS-14 instrument and was operated over a temperature range of 1.8 to 305 K. The samples were cut and lightly (and carefully) polished so as to have parallel surfaces parallel and perpendicular to the $c$-direction with the distances $L$ between the surfaces ranging between approximately $0.3 - 1.4$ mm. Specific heat and thermal expansion data were taken on warming and on the same samples.

Thermodynamic properties of materials are frequently analyzed using the concept of a Gr\"{u}neisen function (or a
Gr\"{u}neisen parameter) \cite{bar99a}. For a single energy scale, $\varepsilon$, the Gr\"{u}neisen
parameter, $\gamma$, is defined as $\gamma = -d \ln \varepsilon/d \ln V$, where $V$ is a molar volume. Using
thermodynamic relations, we can obtain $\gamma(T,V) = \beta V/\chi_S C_p$, where $\beta$ is a volume thermal
expansion coefficient ($\beta = (\partial ln~V/\partial T)_P$), $\chi_S$ is an adiabatic compressibility ($\chi_S
= - (\partial ln~V/\partial P)_S$) and $C_p$ is a heat capacity at a constant pressure.  If, as in many modern materials of interest, more than one contribution to the thermodynamic properties is present (e.g. vibrational, electronic, magnetic, etc.), the Gr\"{u}neisen parameters are not additive, rather the Gr\"{u}neisen parameter for the material is an
average, weighted by the heat capacity contribution of each component \cite{bar99a}: $\gamma = \sum\limits_{r}
\gamma_r C_r/\sum\limits_{r} C_r$. Even with such complexity, the Gr\"{u}neisen parameter behavior in many cases still allows for some qualitative conclusions.

Sometimes, in the analysis of experimental data,
lacking the temperature-dependent compressibility data, the temperature dependence of the Gr\"{u}neisen parameter
can be approximated \cite{pot81a} as being proportional to $\beta/C_p$
under the assumption that the relative temperature dependence of $\chi_S$ is significantly smaller then that of
thermal expansion coefficient or heat capacity. We will follow such approach in this work.
\\

\begin{figure}
\begin{center}
\includegraphics[angle=0,width=120mm]{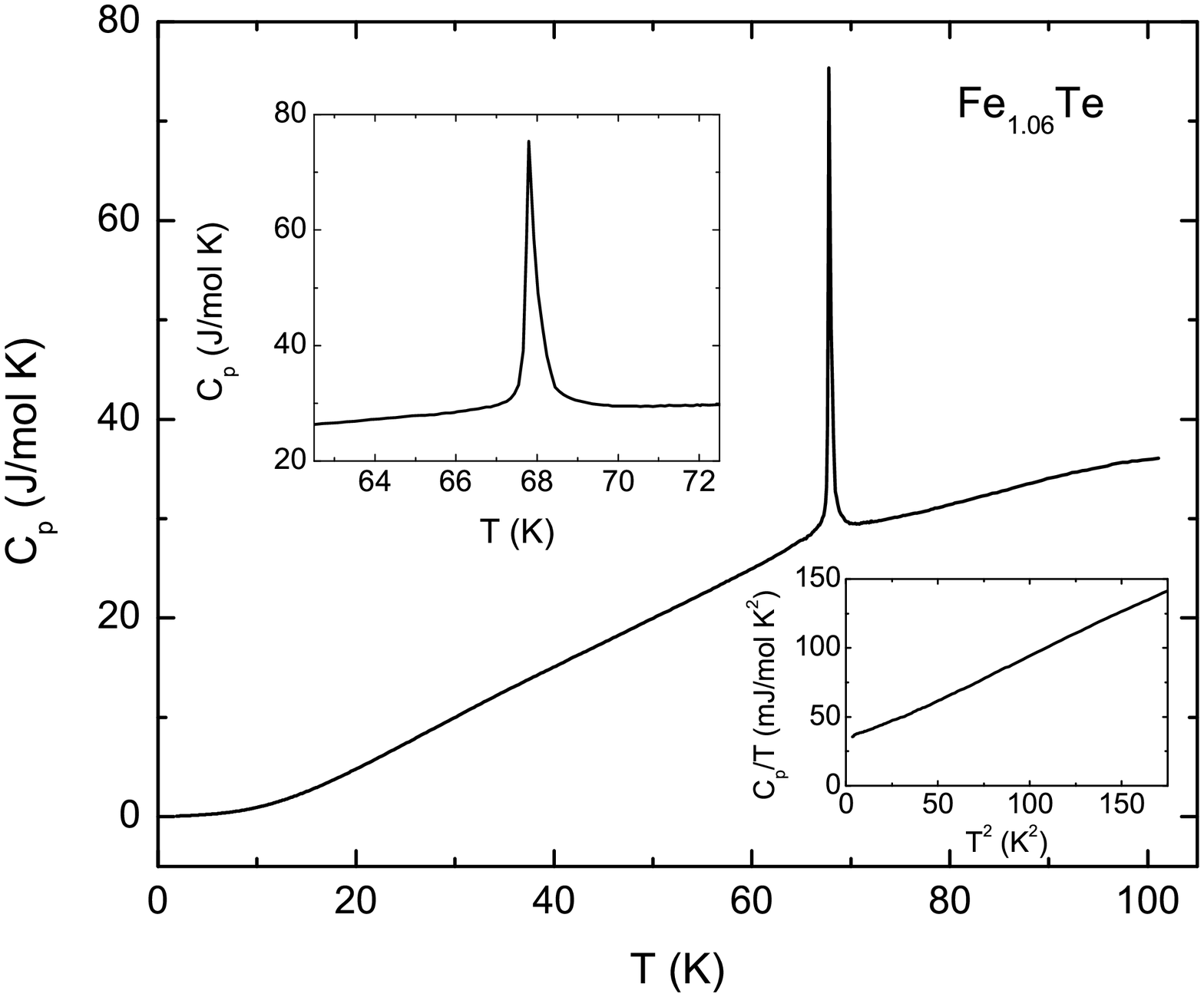}
\end{center}
\caption{Temperature-dependent heat capacity of Fe$_{1.06}$Te single crystal. Left inset: enlarged region near the structural/magnetic phase transition; right inset: low temperature part of the heat capacity plotted as $C_p/T$ vs. $T^2$.}\label{F1}
\end{figure}

\begin{figure}
\begin{center}
\includegraphics[angle=0,width=120mm]{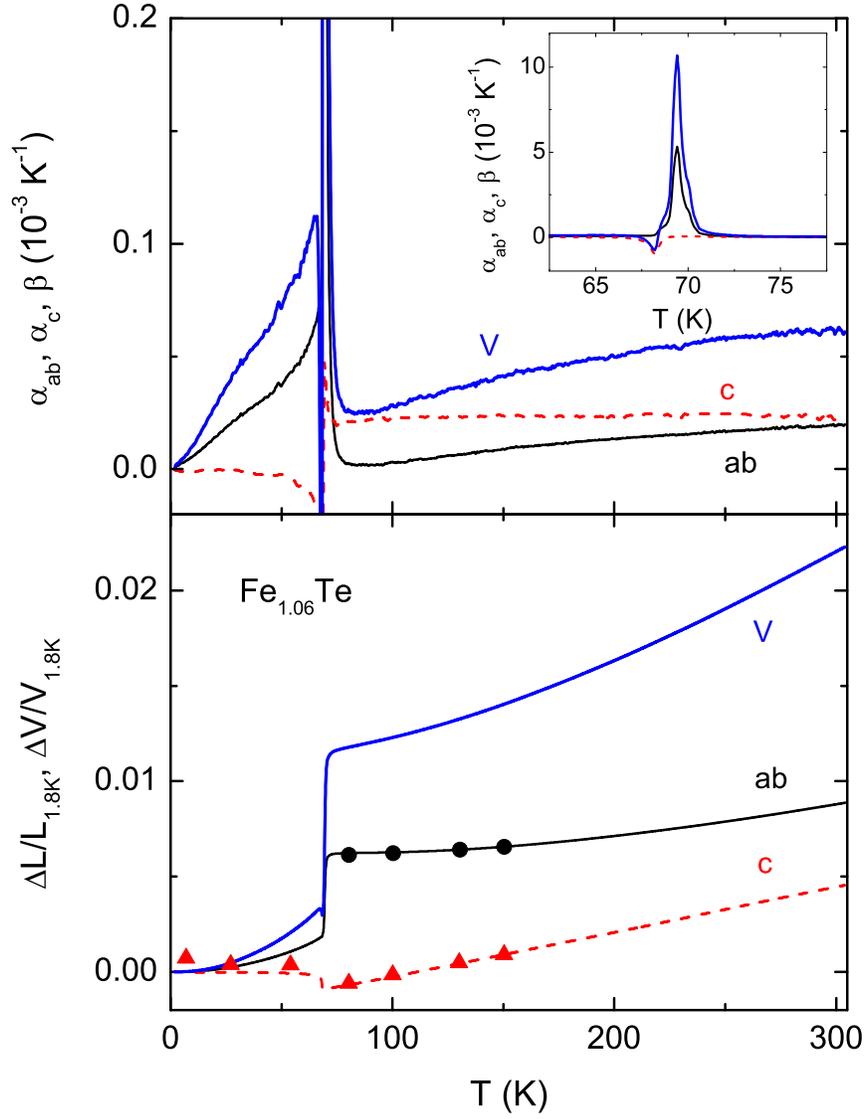}
\end{center}
\caption{(Color online) Anisotropic thermal expansivities (lower panel) and thermal expansion coefficients (upper panel) of Fe$_{1.06}$Te single crystal. Inset to the upper panel: enlarged region near the structural/magnetic phase transition. Symbols: normalized at $T = 150$ K data for Fe$_{1.076}$Te taken from Ref. \onlinecite{bao09a}.}\label{F2}
\end{figure}

The temperature-dependent heat capacity data for the Fe$_{1.06}$Te crystal are shown in Fig. \ref{F1}. A narrow, sharp peak is clearly seen at $\sim 68$ K. The electronic specific heat coefficient is estimated as $\gamma \approx 34$ mJ/mol K$^2$. These data are very similar to those reported for Fe$_{1.05}$Te in Ref. \onlinecite{che09a}. The transition (in Fe$_{1.068}$Te) was identified \cite{lis09a} as being first order, structural and antiferromagnetic. The temperature-depending anisotropic thermal expansivities and thermal expansion coefficients for Fe$_{1.06}$Te are shown in Fig. \ref{F2}. The thermal expansion coefficients are less anisotropic than in BaFe$_2$As$_2$ [\onlinecite{bud09a}] The $c$-axis thermal expansion coefficient is positive and almost temperature independent above the transition and small, negative, weakly temperature-dependent below the transition.  The transition is seen as a sharp feature in each of the measurements. The length in the $ab$ plane decreases, in relative terms, by $\approx 4.4 \cdot 10^{-3}$ on cooling through the transition. The change along the $c$-axis is smaller and of the opposite sign: the relative increase along the $c$-axis is $\approx 9 \cdot 10^{-4}$. We note, however, that the "bulk" thermal expansion measurements yield an average thermal expansion and are not sensitive to possible change in structural symmetry in different phases. Moreover, in the current measurements the exact in-plane direction was not defined. The thermal expansivities are in remarkable agreement with those measured by neutron powder diffraction for Fe$_{1.076}$Te in Ref. \onlinecite{bao09a} (Fig. \ref{F2}). The change at the transition in the $c$-axis is very close to the value reported by neutron scattering (between 80 K and 5 K) \cite{lis09a}, Our in-plane data are within the range of that from neutron scattering, \cite{lis09a} but not the same as reported change in either $a$ or $b$ lattice parameter, that is not surprising considering that our measurements were done for arbitrary in-plane orientation and that there is a possibility of in-plane structural domains below the structural/magnetic transition that will cause some average value to be measured by "bulk" dilatometric techniques.

\begin{figure}
\begin{center}
\includegraphics[angle=0,width=120mm]{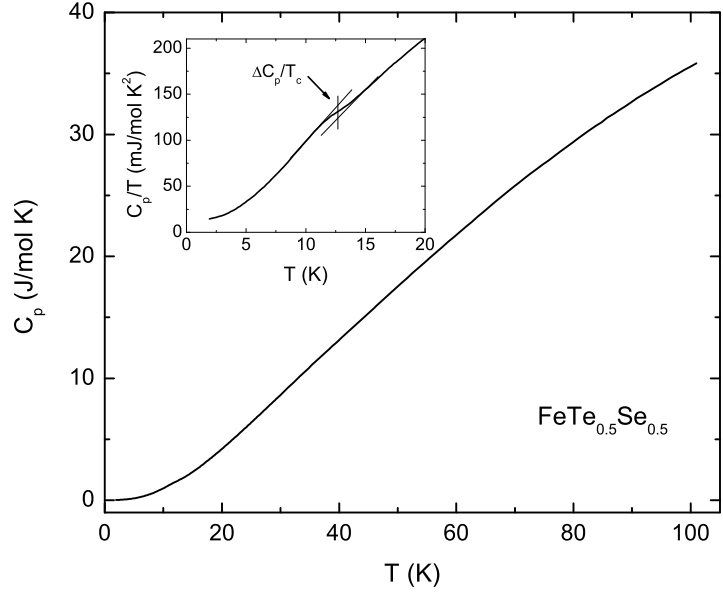}
\end{center}
\caption{Temperature-dependent heat capacity of FeTe$_{0.5}$Se$_{0.5}$ single crystal. Inset: enlarged region near the superconducting transition transition plotted as $C_p/T$ vs. $T$. Lines show how $\Delta C_p/T_c$ value is defined.}\label{F3}
\end{figure}

\begin{figure}
\begin{center}
\includegraphics[angle=0,width=120mm]{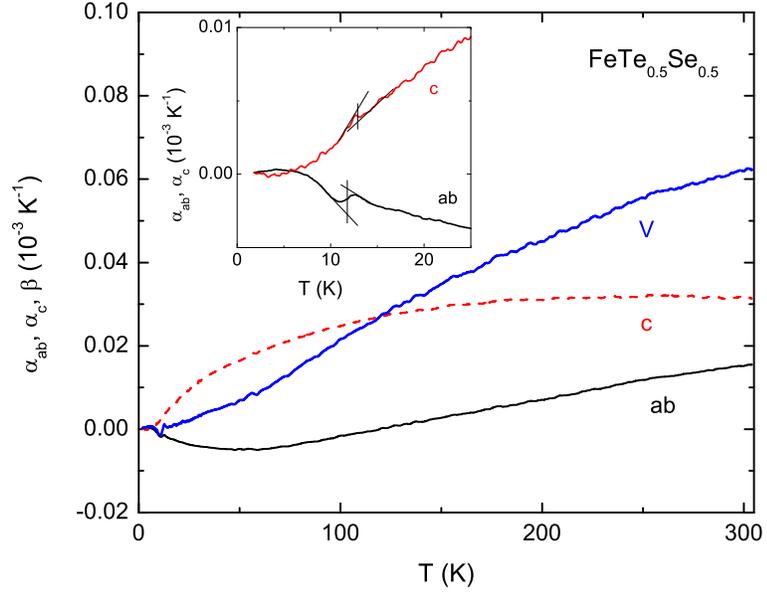}
\end{center}
\caption{(Color online) Anisotropic thermal expansion coefficients  of FeTe$_{0.5}$Se$_{0.5}$ single crystal. Inset: enlarged region near the superconducting transition transition. Lines show how $\Delta \alpha_i$ values are defined.}\label{F4}
\end{figure}

The temperature-dependent heat capacity for FeTe$_{0.5}$Se$_{0.5}$ crystal is shown in Fig. \ref{F3}. A feature associated with a superconducting transition (with an onset $T_c^{onset} \approx 14$ K) is clear in the data. Thermal expansion of the FeTe$_{0.5}$Se$_{0.5}$ crystal (Fig. \ref{F4}) is more anisotropic that that of Fe$_{1.06}$Te. The in-plane thermal expansion is negative below $\sim 120$ K. The features at the superconducting transition are seen in both directions and the changes in the thermal expansion at $T_c$ are of the opposite sign in the in-plane and $c$-axis data sets.

The initial uniaxial pressure derivatives of $T_c$ can be estimated using the Ehrenfest relation for the second order phase transitions \cite{bar99a}:
\begin{displaymath}
dT_c/dp_i = \frac{V_m~\Delta \alpha_i}{\Delta C_p/T_c}
\end{displaymath}
where $V_m$ is the molar volume, $\Delta\alpha_i$ is a change of the linear ($i = ab, c$) thermal expansion coefficient at the superconducting transition, and $\Delta C_p/T_c$ is a change of the specific heat at the superconducting transition divided by $T_c$. Using experimental values: $V_m = 0.26 \cdot 10^{-4}$ m$^3$/mol, \cite{sal09a} $\Delta\alpha_{ab} \approx - 1.8 \cdot 10^{-6}$ K$^{-1}$, $\Delta\alpha_c \approx 0.8 \cdot 10^{-6}$ K$^{-1}$, and $\Delta C_p/T_c \approx 13.4$ mJ/mol K$^2$, we can estimate initial uniaxial pressure derivatives of the superconducting transition temperature in FeTe$_{0.5}$Se$_{0.5}$: $dT_c/dp_{ab} \approx - 0.35$ K/kbar, $dT_c/dp_c \approx 0.16$ K/kbar. This rough estimate of the hydrostatic pressure derivative of $T_c$ is then $dT_c/dP \approx  2 \cdot dT_c/dp_{ab} + dT_c/dp_c \approx - 0.54$ K/kbar. So in-plane pressure should cause a decrease of $T_c$ and pressure along the $c$-axis is expected to cause an increase of $T_c$. The signs of the inferred uniaxial pressure derivatives are the same as for "underdoped" Ba(Fe$_{0.962}$Co$_{0.038}$)$_2$As$_2$, \cite{bud09a} but the absolute values are more moderate; about an order of magnitude smaller.

\begin{figure}
\begin{center}
\includegraphics[angle=0,width=120mm]{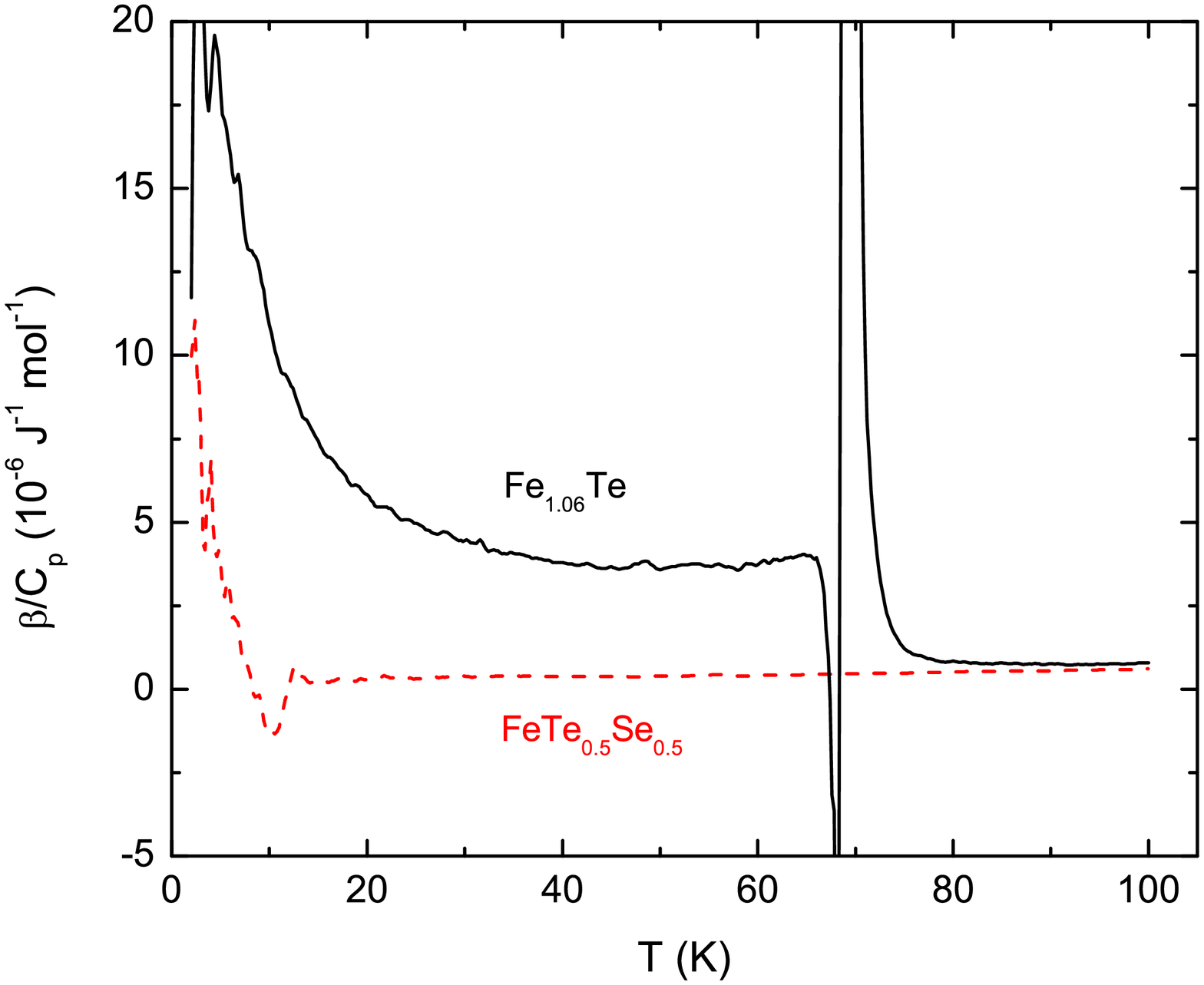}
\end{center}
\caption{(Color online) Gr\"uneisen parameters, $\beta/C_p$, of Fe$_{1.06}$Te and FeTe$_{0.5}$Se$_{0.5}$ single crystal.}\label{F5}
\end{figure}

The temperature-dependent Gr\"uneisen parameters, in the form of $\beta/C_p$, (volume thermal expansion, $\beta$, is defined here as $\beta = 2 \cdot \alpha_{ab} + \alpha_c$) for Fe$_{1.06}$Te and FeTe$_{0.5}$Se$_{0.5}$ are shown in Fig. \ref{F5}. (The excessive noise below $\sim 5$ K could be caused by the division by small $C_p$ values.) For FeTe$_{0.5}$Se$_{0.5}$ the $\beta/C_p$ is practically temperature-independent above $\sim 15$ K. For Fe$_{1.06}$Te the value of $\beta/C_p$ at temperatures above the structural/magnetic phase transition is very close to that of FeTe$_{0.5}$Se$_{0.5}$, however below the transition the  Gr\"uneisen parameter of Fe$_{1.06}$Te is significantly higher, and, close to the transition is only weakly temperature-dependent. Most probably this difference is due to the magnetic contribution (the heat capacity and the inferred Debye temperature are continuous if the region around transition is excluded, so probably the change in the phonon term is not so drastic through the transition) however other contributions cannot be excluded and more studies are required to clarify this issue.
\\

In summary, thermal expansion of  Fe$_{1.06}$Te and FeTe$_{0.5}$Se$_{0.5}$ is anisotropic, phase transitions are clearly seen. The signs of the inferred uniaxial pressure derivatives of $T_c$ in FeTe$_{0.5}$Se$_{0.5}$ are opposite for in-plain ($dT_c/dp_{ab} < 0$) and $c$-axis ($dT_c/dp_c > 0$) pressures. The Gr\"uneisen parameters for both materials are similar and only weakly temperature-dependent above $\sim 80$ K. At low temperatures (in the magnetically ordered phase) the magnetic contribution to the Gr\"uneisen parameter in Fe$_{1.06}$Te appears to be significantly larger than electron and phonon contributions combined.

\begin{acknowledgments}

Work at the Ames Laboratory was supported by the US Department of Energy - Basic Energy Sciences under Contract No. DE-AC02-07CH11358. Research at Oak Ridge National Laboratory was sponsored by the Division of Materials Sciences and Engineering, Office of Basic Energy Sciences, US Department of Energy. We are indebted to George M. Schmiedeshoff for his help in establishing dilatometry technique in Ames Laboratory Novel Materials and Ground States Group and for many propitious advices. M. T. C. Apoo is gratefully acknowledged.

\end{acknowledgments}

\clearpage

\clearpage

\clearpage

\clearpage

\clearpage

\begin{thebibliography}{99}

\bibitem{kam08a} Y. Kamihara, T. Watanabe, M. Hirano, H. Hosono, Journal of the American Chemical Society  {\bf 130}, 3296 (2008).

\bibitem{rot08a} Marianne Rotter, Marcus Tegel, and Dirk Johrendt, Phys. Rev. Lett. {\bf 101}, 107006 (2008).

\bibitem{wan08a} X. C. Wang, Q. Q. Liu, Y. X. Lv, W. B. Gao, L. X. Yang, R. C. Yu, F. Y. Li, C. Q. Jin, Solid State Comm. {\bf 148}, 538 (2008).

\bibitem{hsu08a} Fong-Chi Hsu, Jiu-Yong Luo, Kuo-Wei Yeh, Ta-Kun Chen, Tzu-Wen Huang, Phillip M. Wu, Yong-Chi Lee,
Yi-Lin Huang, Yan-Yi Chu, Der-Chung Yan, and Maw-Kuen Wu, Proc. Natl. Acad. Sci. USA {\bf 105}, 14262 (2008).

\bibitem{miz08a} Yoshikazu Mizuguchi, Fumiaki Tomioka, Shunsuke Tsuda, Takahide Yamaguchi,
and Yoshihiko Takano, arXiv:0811.1123v3, unpublished.

\bibitem{sal09a} B. C. Sales, A. S. Sefat, M. A. McGuire, R. Y. Jin, D. Mandrus, and Y. Mozharivskyj, Phys. Rev. B {\bf 79}, 094521 (2009).

\bibitem{kro98a} F. Kromer, R. Helfrich, M. Lang, F. Steglich, C. Langhammer, A. Bach, T. Michels, J. S. Kim, and G. R. Stewart, Phys. Rev. Lett., {\bf 81} 4476 (1998).

\bibitem{bud09a} S. L. Bud'ko, N. Ni, S. Nandi, G. M. Schmiedeshoff, and P. C. Canfield, Phys. Rev. B {\bf 79}, 054525 (2009).

\bibitem{har09a} Fr\'ed\'eric Hardy, Peter Adelmann, Thomas Wolf, Hilbert v. L\"ohneysen, and Christoph Meingast, Phys. Rev. Lett. {\bf 102}, 187004 (2009).

\bibitem{luz09a} M. S. da Luz, J. J. Neumeier, R. K. Bollinger, A. S. Sefat, M. A. McGuire, R. Jin, B. C. Sales, and D. Mandrus, Phys. Rev. B {\bf 79}, 214505 (2009).

\bibitem{bar99a} T. H. K. Barron and G. K. White, {\it Heat Capacity and Thermal Expansion at Low Temperatures} (Kluwer Academic/Plenum, New York, 1999).

\bibitem{sch06a} G. M. Schmiedeshoff, A. W. Lounsbury, D. J. Luna, S. J. Tracy, A. J. Schramm, S. W. Tozer, V. F. Correa, S. T. Hannahs,
T. P. Murphy, E. C. Palm, A. H. Lacerda, S. L. Bud'ko, P. C. Canfield, J. L. Smith, J. C. Lashley, and J. C. Cooley,  Rev.
Sci. Instrum. {\bf 77}, 123907 (2006).

\bibitem{pot81a} R. Pott, R. Schefzyk, D. Wohlleben, and A. Junod, Z. Phys. B {\bf 44}, 17 (1981).

\bibitem{che09a} G. F. Chen, Z. G. Chen, J. Dong, W. Z. Hu, G. Li, X. D. Zhang, P. Zheng, J. L. Luo, and N. L. Wang, Phys. Rev. B {\bf 79}, 140509 (2009).

\bibitem{bao09a} Wei Bao, Y. Qiu, Q. Huang, M. A. Green, P. Zajdel, M. R. Fitzsimmons, M. Zhernenkov, S. Chang, Minghu Fang, B. Qian, E. K. Vehstedt, Jinhu Yang, H. M. Pham, L. Spinu, and Z. Q. Mao, Phys. Rev. Lett. {\bf 102}, 247001 (2009).

\bibitem{lis09a} Shiliang Li, Clarina de la Cruz, Q. Huang, Y. Chen, J. W. Lynn, Jiangping Hu, Yi-Lin Huang, Fong-Chi Hsu,
Kuo-Wei Yeh, Maw-Kuen Wu, and Pengcheng Dai, Phys. Rev. B {\bf 79}, 054503 (2009).


\end{thebibliography}
\end{document}